\documentstyle[amssymb,prb,aps,epsf]{revtex}

\begin{document}
\title{Conductance modulation by spin precession in non-collinear \\
ferromagnet-normal metal-ferromagnet systems.}
\author{Daniel Huertas Hernando, Yu. V. Nazarov, Arne Brataas$\dagger $ and Gerrit
E.W. Bauer}
\address{Department of Applied Physics and Delft Institute of Microelectronics and\\
Submicrontechnology, \\
Delft University of Technology, Lorentzweg 1, 2628 CJ Delft, The Netherlands%
\\
$\dagger $Harvard University, Lyman Laboratory of Physics, Cambridge MA\\
02138, USA}
\date{\today}
\maketitle

\begin{abstract}
We study diffusive transport through ferromagnet - normal metal -
ferromagnet (F-N-F) systems, with arbitrary but fixed magnetization
directions of the ferromagnetic reservoirs and orientations of a magnetic
field applied to the normal metal. For non-collinear configurations, the
complex mixing conductance $G^{\uparrow \downarrow }$ describes the
transport of spins non-collinear to the magnetizations of the ferromagnetic
reservoirs. When $%
\mathop{\rm Im}%
G^{\uparrow \downarrow }\neq 0$, the total conductance of the system in the
presence of a magnetic field can be asymmetric with respect to time
reversal. The total conductance changes non-monotonically with the magnetic
field strength for different magnetic configurations. This modulation of the
conductance is due to the precession of the spin accumulation in the normal
metal. The difference between the conductance of the parallel and
antiparallel configurations can be either positive or negative as a function
of the applied magnetic field. This effect should be best observable on Al
single crystals attached to ferromagnetic electrodes by means of tunnel
junctions or metallic contacts.
\end{abstract}


\section{\bf INTRODUCTION}

\qquad\ In hybrid systems of ferromagnetic and normal metals, interesting
phenomena can appear due the interplay between charge and spin. The
discovery of the giant magnetoresistance (GMR) effect in metallic magnetic
multilayers,\cite{RefBaibich} has motivated a large number of studies on the
transport properties of such systems.\cite{Ref Camley} The GMR is caused by
spin dependent scattering in the system. Most studies concentrated on
collinear configurations (parallel and antiparallel configurations). There
are several papers which cover non-collinear magnetizations, both theoretical
\cite{RefSlonczewski} and experimental.\cite{RefDauguet}

Magnetoelectronic multiterminal devices reveal new physics,\cite{RefMJohnson}
but may also lead to novel applications, {\em e.g.} non-volatile
electronics. Johnson and Silsbee investigated spin dependent effects in a
3-terminal device.\cite{RefMJohnson} They found transistor effects that
depend on the relative orientation of the magnetization of the ferromagnets. 
\cite{RefJohnsontrans} More recently, a ferromagnetic single-electron
transistor in a three terminal configuration has been realized\cite{RefOno}
and studied theoretically.\cite{RefBarnas} In this case the source-drain
current also depends on the relative orientation of the magnetizations.

Brataas {\em et al.}\cite{RefCircuit} give a unified semiclassical picture
for electron and spin transport in such systems. Their formalism is inspired
by the circuit theory of the Andreev reflection,\cite{RefYuli} and is
applicable to systems with non-collinear magnetization directions and an
arbitrary number and variety of contacts between the ferromagnetic and the
normal metals. However, the simple circuit theory of Ref. %
\onlinecite{RefCircuit} only holds when the resistances of the contacts
between the ferromagnetic and the normal metals are much higher than the
resistance of the normal metal itself, thus fails when the size $L$ of the
system in the transport direction becomes too large. Moreover, when the size
of the system is larger than the spin diffusion length ($L\gg l_{sf}$), the
presence of spin-diffusion in the normal metal requires a more complicated
description with spatially dependent spin distribution functions.

In the present paper, we present a study of the transport properties of
simple F-N-F systems (see Fig. 1), taking into account different
magnetizations of the ferromagnetic reservoirs and spin-diffusion in the
normal metal. At low temperatures, spin-flip can be due to spin-orbit
interactions and scattering by defects or impurities. Exchange scattering by
paramagnetic impurities also flips the spin (see {\em e.g.} Appendix A in
Ref.\onlinecite{RefValet}). The length of the normal metal $L$ is assumed to
be much larger than the mean-free path $l_{f}$, so electronic transport may
be described by the diffusion equation. On the other hand, we allow the spin
diffusion length $l_{sf}$, which is the length scale on which an electron
looses its spin in diffusive transport, to be much smaller, of the same
order, or much larger than the size of the system $L$. Under an applied
bias, ferromagnetic reservoirs inject a spin-current, causing a
non-equilibrium magnetization or ``spin accumulation'' in the normal metal.
We are interested in the different mechanisms that reduce and also rotate
this spin accumulation. For non-collinear configurations the physics of spin
injection is more subtle than in the simple collinear case, since it
requires generalized boundary conditions for transport through a single
ferromagnetic-normal metal (F-N) contact.\cite{RefCircuit} In general, such
a contact is charaterized not only by the conventional spin dependent
conductances $G^{\uparrow },G^{\downarrow }$, which describe the transport
of spins collinear to the magnetization of the ferromagnetic reservoir, but
also by the $\left( \text{complex}\right) $ {\em mixing }conductance $%
G^{\uparrow \downarrow }$ (see Ref.\onlinecite{RefCircuit}), that contains
information about the transport of spins oriented perpendicular to the
magnetization of the ferromagnetic reservoir.

We are also interested in the effect of a magnetic field applied to the
diffusive normal metal in arbitrary directions. In this case we assume that
the magnetic field only couples to the spin degrees of freedom. Our approach
is similar to the treatment of a precessing magnetic field applied to a
diffusive metal in Ref.\onlinecite{Ref vanLangen}.

In section II we introduce and solve the basic equations for the diffusive
spin transport, showing the general expression for the non-equilibrium
distribution function in the normal metal. In section III we discuss the
boundary conditions of the problem. In section IV we obtain analytical
expressions for the total conductance of the system in collinear
configurations and in the absence of applied magnetic field. We also obtain
analytical expressions for the total conductance in the case of
non-collinear magnetization directions, zero magnetic field and no spin-flip
scattering. In section V we calculate numerically the conductance in the
general case. In section VI we summarize and discuss our results.

\section{\bf DIFFUSIVE SPIN TRANSPORT}

When a bias is applied to our F-N-F device, a spin current is injected from
the ferromagnetic reservoirs into the normal metal, causing a
non-equilibrium magnetization or spin accumulation. For an arbitrary
magnetic configuration of the system, the spin accumulated in the normal
metal can be oriented in differents directions. If we take the spin
quantization axis parallel to the magnetization of one of the ferromagnetic
reservoirs, we need to take into account spins oriented perpendicular to
this quantization axis, which can be described as a superposition of up ($%
\uparrow $) and down ($\downarrow $) spin states. We study a geometry
invariant to translations in the lateral direction, so all quantities depend
only on one spatial coordinate $(x)$. The spin-polarized electron
distribution is characterized by a $2\times 2$ matrix in spin space of the
form: 
\begin{equation}
\hat{f}^{N}(x)=\left( 
\begin{array}{cc}
\begin{array}{c}
f_{\uparrow \uparrow }^{N}(x)
\end{array}
& 
\begin{array}{c}
f_{\uparrow \downarrow }^{N}(x)
\end{array}
\\ 
\begin{array}{c}
\\ 
f_{\downarrow \uparrow }^{N}(x)
\end{array}
& 
\begin{array}{c}
\\ 
f_{\downarrow \downarrow }^{N}(x)
\end{array}
\end{array}
\right) .  \label{distribfN.}
\end{equation}
When the size of the system $L$ is larger than the spin diffusion length $%
l_{sf}$, $\hat{f}^{N}(x)$ depends on the position. Here we are interested in
transport under the condition $l_{f}\ll l_{sf}$, where $l_{f}=v_{F}\left(
1/\tau +1/\tau _{sf}\right) ^{-1}$ is the{\em \ mean free path}, $v_{F}$ is
the Fermi velocity, $\tau $ the spin-conserving scattering time and $\tau
_{sf}$ the spin-flip scattering time. Both $\tau $ and $\tau _{sf}$ are
considered isotropic in momentum space$.$ The spin diffusion length $l_{sf}$
is defined as $l_{sf}=\sqrt{D\tau _{sf}}$, where $D=v_{F}l_{f}/d$, is the
spin-independent diffusion coefficient of the normal metal ($d=1,2,3$ is the
dimension of the normal metal). So under the condition $l_{f}\ll l_{sf}$, we
obtain for diffusive spin transport in the steady state the following $%
2\times 2$ matrix equations for $\hat{f}^{N}(x)$

\begin{eqnarray}
D\frac{\partial ^{2}\hat{f}^{N}(x)}{\partial x^{2}} &=&\frac{1}{\tau _{sf}}%
\left( \hat{f}^{N}(x)-{\bf \hat{1}}\frac{\mbox{Tr}\left( \hat{f}%
^{N}(x)\right) }{2}\right)  \label{eqspinrelax} \\
\hat{\jmath}^{N}(x) &=&-D\frac{\partial \hat{f}^{N}(x)}{\partial x}.
\label{currentdens.}
\end{eqnarray}
where ${\bf \hat{1}}$ is the unit matrix and where the electron charge $e$
is assumed to be equal to one. Eq. (\ref{eqspinrelax}) describes the
relaxation of the spin accumulation due to spin-flip scattering, and (\ref
{currentdens.}) relates the current density matrix $\hat{\jmath}^{N}(x)$ and 
$\hat{f}^{N}(x)$. In the case of collinear transport, our matrix equations
simply reduce to

\begin{eqnarray}
\frac{\partial ^{2}f_{s}^{N}(x)}{\partial x^{2}} &=&\frac{1}{2}\frac{%
f_{s}^{N}(x)-f_{-s}^{N}(x)}{l_{sf}^{2}}\text{ \ \ }  \label{valetfert1} \\
j_{s}^{N}(x) &=&-D\frac{\partial f_{s}^{N}(x)}{\partial x}\text{ \ \ \ \ }
\label{valetfert2}
\end{eqnarray}
where $s=\left( \uparrow ,\downarrow \right) $. Eqs. (\ref{valetfert1}) and (%
\ref{valetfert2}) have been extensively used for collinear transport in F-N
multilayers in which the current flows perpendicular to the planes of the
interfaces (CPP geometry). \cite{RefValet,RefSonKem}

We are also interested in the effect of an external magnetic field applied
to the normal metal in an arbitrary direction. We know that the magnetic
Zeeman energy associated with the coupling between the magnetic field and
the spin of the electrons is given by $g\mu _{B}{\bf \hat{\sigma}\cdot }\vec{%
B}/2$, where $\mu _{B}$ is the Bohr magneton, $g$ is the gyromagnetic ratio, 
${\bf \hat{\sigma}=}\left( \hat{\sigma}_{x},\text{ }\hat{\sigma}_{y},\text{ }%
\hat{\sigma}_{z}\right) $ is the vector of Pauli matrices and $\vec{B}$ is
the external magnetic field. Semiclassically, we can write for the spin
dynamics (see {\em e.g.} Ref.\onlinecite{RefRammer}) 
\begin{equation}
\frac{\partial \text{ }\hat{f}^{N}(x)}{\partial t}=\frac{i}{\hslash }\left[ 
\frac{g\mu _{B}}{2}\left( {\bf \hat{\sigma}\cdot }\vec{B}\right) ,\text{ }%
\hat{f}^{N}(x)\right] _{-}.  \label{comutf}
\end{equation}
Then, in the steady state: 
\begin{equation}
D\frac{\partial ^{2}\hat{f}^{N}(x)}{\partial x^{2}}=\frac{1}{\tau _{sf}}%
\left( \hat{f}^{N}(x)-{\bf \hat{1}}\frac{\mbox{Tr}\left( \hat{f}%
^{N}(x)\right) }{2}\right) -\frac{i}{\hslash }\left[ \frac{g\mu _{B}}{2}%
\left( {\bf \hat{\sigma}\cdot }\vec{B}\right) ,\text{ }\hat{f}^{N}(x)\right]
_{-}.  \label{eqcompmagnet}
\end{equation}
Using the properties of the Pauli matrices we can express the
non-equilibrium distribution matrix $\hat{f}^{N}(x)$ as: 
\begin{equation}
\hat{f}^{N}(x)=f_{0}(x){\bf \hat{1}}+{\bf \hat{\sigma}\cdot }\vec{f}(x)
\label{transpaulimat}
\end{equation}
where $f_{0}(x)$ is a scalar and $\vec{f}(x)=\left(
f_{x}(x),f_{y}(x),f_{z}(x)\right) $ is a three component vector. $f_{0}(x)$
is the particle or {\em spin-independent }distribution function. On the
other hand, $f_{z}(x)$ describes the ``spin polarization'' on the system,
and $f_{x}(x)$ and $f_{y}(x)$ contain information about the spins oriented
perpendicular to the quantization axis. We call the three component vector $%
\vec{f}(x)$ the {\em spin-dependent }distribution{\em \ }function. Using (%
\ref{transpaulimat}), we separate (\ref{eqcompmagnet}) into two
contributions, one for the {\em spin-independent }part and another for the 
{\em spin-dependent} part:

\begin{mathletters}%
%

\begin{eqnarray}
\frac{\partial ^{2}f_{0}(x)}{\partial x^{2}} &=&0  \label{finaleqs1} \\
\frac{\partial ^{2}\vec{f}(x)}{\partial x^{2}} &=&\frac{1}{l_{sf}^{2}}\vec{f}%
(x)+\left( \frac{g\mu _{B}}{\hslash }\frac{\vec{B}}{D}\times \text{ }\vec{f}%
(x)\right) .  \label{finaleqs3}
\end{eqnarray}
\label{finaleqs}%
\end{mathletters}%
%
The spin-independent part (Eq.(\ref{finaleqs1})), is the conventional result
for diffusive particle transport. Similar to Eq. (\ref{valetfert2}), the 
{\em particle} current density $j_{0}^{N}(x)$ reads 
\begin{equation}
\text{ }j_{0}^{N}(x)=-D\frac{\partial \left[ \mbox{Tr}\left( \hat{f}%
^{N}(x)\right) \right] }{\partial x}=-2D\frac{\partial f_{0}(x)}{\partial x}.
\label{partcurrentdens.}
\end{equation}
Eqs. (\ref{finaleqs1}) and (\ref{partcurrentdens.}) express the particle
current conservation 
\[
\frac{\partial j_{0}^{N}(x)}{\partial x}=0. 
\]
The general solution of Eq. (\ref{finaleqs1}) is 
\begin{equation}
f_{0}(x)={\cal P}+{\cal O}x.  \label{gensol1}
\end{equation}
Eq. (\ref{finaleqs3}) describes how the spin accumulation relaxes by
spin-flip scattering and by the spin {\em precession} around the magnetic
field. This equation can be written in a general matrix form as: 
\[
\frac{\partial ^{2}\vec{f}(x)}{\partial x^{2}}={\bf A}\vec{f}(x). 
\]
The eigengenvalues associated with the matrix ${\bf A}$ are: 
\begin{eqnarray*}
\lambda _{o} &=&\frac{1}{l_{sf}^{2}} \\
\lambda _{+} &=&\frac{1}{l_{sf}^{2}}+i\left| \vec{h}\right| \\
\lambda _{-} &=&\frac{1}{l_{sf}^{2}}-i\left| \vec{h}\right| .
\end{eqnarray*}
where we have introduced the vector $\vec{h}=g\mu _{B}\vec{B}/\hslash D$,
which describes the ``effectiveness'' of the magnetic field in a diffusive
metal. The eigenvector associated with $\lambda _{o}$ is 
\[
\vec{v}_{o}=\frac{1}{\left| \vec{h}\right| }\left( 
\begin{array}{c}
h_{x} \\ 
h_{y} \\ 
h_{z}
\end{array}
\right) . 
\]
On the other hand, $\lambda _{+}$ and $\lambda _{-}$ have associated two
complex conjugated eigenvectors $\vec{v}_{+}=\vec{v}_{1}+i\vec{v}_{2}$ and $%
\vec{v}_{-}=\vec{v}_{1}-i\vec{v}_{2}$, where 
\[
\vec{v}_{1}=\frac{1}{\sqrt{\left( h_{x}^{2}+h_{y}^{2}\right) }\left| \vec{h}%
\right| }\left( 
\begin{array}{c}
h_{x}h_{z} \\ 
h_{y}h_{z} \\ 
-\left( h_{z}^{2}+h_{x}^{2}\right)
\end{array}
\right) 
\]
and 
\[
\text{ }\vec{v}_{2}=\frac{1}{\sqrt{h_{x}^{2}+h_{y}^{2}}}\left( 
\begin{array}{c}
h_{y} \\ 
-h_{x} \\ 
0
\end{array}
\right) . 
\]
The general solution of Eq. (\ref{finaleqs3}), can then be written in terms
of the eigenvalues $\lambda _{o},$ $\lambda _{+},$ $\lambda _{-}$ and
vectors $\vec{v}_{o},$ $\vec{v}_{1},$ $\vec{v}_{2}$ as:

\begin{equation}
\vec{f}(x)=\left\{ 
\begin{array}{c}
{\cal A}\stackrel{}{\vec{v}_{o}}\cosh (x/l_{sf})+{\cal B}\stackrel{}{\vec{v}%
_{o}}\sinh (x/l_{sf}) \\ 
+{\cal C}\text{ \ }\left[ \vec{v}_{1}\cosh (X)\cos (Y)-\vec{v}_{2}\sinh
(X)\sin (Y)\right] \\ 
-{\cal D}\text{ }\left[ \vec{v}_{1}\sinh (X)\sin (Y)+\vec{v}_{2}\cosh
(X)\cos (Y)\right] \\ 
+{\cal E}\text{ \ }\left[ \vec{v}_{1}\sinh (X)\cos (Y)-\vec{v}_{2}\cosh
(X)\sin (Y)\right] \\ 
-{\cal F}\text{ \ }\left[ \vec{v}_{1}\cosh (X)\sin (Y)+\vec{v}_{2}\sinh
(X)\cos (Y)\right] .
\end{array}
\right\}
\end{equation}
where 
\begin{eqnarray*}
X &=&\sqrt{\frac{1+\sqrt{1+\alpha ^{2}}}{2}}\frac{x}{l_{sf}} \\
Y &=&\sqrt{\frac{-1+\sqrt{1+\alpha ^{2}}}{2}}\frac{x}{l_{sf}}.
\end{eqnarray*}
and where the dimensionless constant $\alpha =g\omega _{L}\tau _{sf}$ $%
=\left| \vec{h}\right| l_{sf}^{2}$ is the ratio between spin-flip and
precession relaxation mechanisms. $\omega _{L}=\mu _{B}\left| \vec{B}\right|
/\hslash $ is the (Larmor) frequency for the spin precession around the
magnetic field.\cite{comment1} The solution associated with $\lambda _{o}$
describes the relaxation of the spin accumulation due to spin-flip
scattering, and the two complex conjugated solutions associated with $%
\lambda _{+}$ and $\lambda _{-}$ describe the relaxation and precession of
the spins due to the coupling with the magnetic field. The eight{\bf \ }real
constants (${\cal O},{\cal P},{\cal A},{\cal B},{\cal C},{\cal D},{\cal E},%
{\cal F}$) must be determined by the boundary conditions.

\section{\bf BOUNDARY CONDITIONS}

We consider two ferromagnetic reservoirs attached to a diffusive normal
metal through some arbitrary contacts, as shown in Fig. 1. The ferromagnetic
reservoirs are supposed to be large and in local equilibrium at chemical
potentials $\mu _{{\cal L},{\cal R}}$ (${\cal L}$, ${\cal R}$ denotes left
and right reservoir respectively), and with energy-dependent diagonal
distribution matrices in spin space $\hat{f}_{{\cal L},{\cal R}%
}^{F}(\epsilon )$. The components of $\ \hat{f}_{{\cal L},{\cal R}%
}^{F}(\epsilon )$ are given by the Fermi-Dirac distribution function $%
f^{FD}(\epsilon ,\mu _{{\cal L},{\cal R}})$, and the direction of the
magnetization in each ferromagnetic reservoir is denoted by the unit vector $%
{\bf \vec{m}}_{{\cal L},{\cal R}}$. The current through the system and the
non-equilibrium distribution function in the normal metal are completely
determined by the relative orientation of the magnetization directions in
the ferromagnetic reservoirs, the contact conductances, the normal metal
conductance, the spin diffusion length and the magnetic field.

The current through an F-N contact is given in Ref.\onlinecite{RefCircuit}
in terms of the microscopic scattering matrices of the Landauer-B\"{u}ttiker
formalism.\cite{RefButtiker} According to Eq. (3) of Ref.%
\onlinecite{RefCircuit}, the particle current through a single contact
directed into the normal metal can be written as 
\begin{eqnarray}
\hat{\imath}^{C}(x) &=&G^{\uparrow }\hat{u}^{\uparrow }\left( \hat{f}^{F}-%
\hat{f}^{N}(x)\right) \hat{u}^{\uparrow }+G^{\downarrow }\hat{u}^{\downarrow
}\left( \hat{f}^{F}-\hat{f}^{N}(x)\right) \hat{u}^{\downarrow }
\label{contactcurrent} \\
&&-G^{\uparrow \downarrow }\hat{u}^{\uparrow }\hat{f}^{N}(x)\hat{u}%
^{\downarrow }-\left( G^{\uparrow \downarrow }\right) ^{\ast }\hat{u}%
^{\downarrow }\hat{f}^{N}(x)\hat{u}^{\uparrow }  \nonumber
\end{eqnarray}
where $\hat{f}^{N}(x)$ and $\hat{f}_{{\cal L},{\cal R}}^{F}$ are isotropic
distribution functions, $G^{\uparrow }$ and $G^{\downarrow }$ are the
conventional spin-dependent conductances, which describe the transport of
spins oriented in the direction of the magnetization of the adjacent{\em \ }%
ferromagnetic reservoir, and $G^{\uparrow \downarrow }=%
\mathop{\rm Re}%
G^{\uparrow \downarrow }+i%
\mathop{\rm Im}%
G^{\uparrow \downarrow }$ is the {\em mixing conductance,} which contains
information about the transport of spins oriented in perpendicular direction
to the magnetization of the ferromagnetic reservoir. The matrices $\hat{u}%
^{\uparrow }=\left( {\bf \hat{1}+}\text{ }{\bf \hat{\sigma}}\cdot {\bf \vec{m%
}}\right) /2,$ and $\hat{u}^{\downarrow }=\left( {\bf \hat{1}-}\text{ }{\bf 
\hat{\sigma}}\cdot {\bf \vec{m}}\right) /2$ define the basis in which the
spin-quantization axis is parallel to the magnetization of the ferromagnet
(for details see Ref.\onlinecite{RefArne}). Eq. (\ref{contactcurrent})
relates the spin current through the contact $\hat{\imath}^{C}(x)$ and the
non-equilibrium distribution matrix $\hat{f}^{N}(x)$ in the normal metal.
Due to current conservation, Eq. (\ref{contactcurrent}) is equal, at each
contact, to the particle current per energy interval in the normal metal
(see Fig. 2). The particle current per energy interval is related with the
current density $\hat{\jmath}^{N}(x)$ as, $\hat{\imath}^{N}(x)=S$ $\nu
_{_{DOS}}$ $\hat{\jmath}^{N}(x)$, where $S$ is the surface perpendicular to
the transport direction and $\nu _{_{DOS}}$ is the density of states of the
normal metal. Using Eq. (\ref{currentdens.}) $\hat{\imath}^{N}(x)$ is 
\begin{equation}
\hat{\imath}^{N}(x)=-S\nu _{_{DOS}}D\frac{\partial \hat{f}^{N}(x)}{\partial x%
}.  \label{currentNORMAL}
\end{equation}
So we have 
\begin{mathletters}%
%
\begin{equation}
\hat{\imath}^{C}(x=0^{+})\text{ }=\text{ }\hat{\imath}^{N}(x=0^{+})
\label{boundary1}
\end{equation}
for the left contact ($x=0^{+}$) and 
\begin{equation}
\hat{\imath}^{N}(x=L^{-})\text{ }=\text{ }\hat{\imath}^{C}(x=L^{-})
\label{boundary2}
\end{equation}
for the right contact ($x=L^{-}$).%
\end{mathletters}%
%
\ By substituting (\ref{contactcurrent}) and (\ref{currentNORMAL}) into (\ref
{boundary1}) and (\ref{boundary2}), we obtain the boundary conditions for
the left contact ($x=0^{+}$):%
\begin{mathletters}%
%
\begin{eqnarray}
&&-S\text{ }\nu _{_{DOS}}\text{ }D\frac{\partial \hat{f}^{N}(x)}{\partial x}%
_{x=0^{+}}+  \label{system1} \\
&&G^{\uparrow }\hat{u}^{\uparrow }\hat{f}^{N}(0^{+})\hat{u}^{\uparrow
}+G^{\downarrow }\hat{u}^{\downarrow }\hat{f}^{N}(0^{+})\hat{u}^{\downarrow
}+  \nonumber \\
&&G^{\uparrow \downarrow }\hat{u}^{\uparrow }\hat{f}^{N}(0^{+})\hat{u}%
^{\downarrow }+\left( G^{\uparrow \downarrow }\right) ^{\ast }\hat{u}%
^{\downarrow }\hat{f}^{N}(0^{+})\hat{u}^{\uparrow }  \nonumber \\
&=&G^{\uparrow }\hat{u}^{\uparrow }\hat{f}_{{\cal L}}^{F}\hat{u}^{\uparrow
}+G^{\downarrow }\hat{u}^{\downarrow }\hat{f}_{{\cal L}}^{F}\hat{u}%
^{\downarrow }  \nonumber
\end{eqnarray}
and for the right contact ($x=L^{-}$): \qquad \qquad 
\begin{eqnarray}
&&S\text{ }\nu _{_{DOS}}\text{ }D\frac{\partial \hat{f}^{N}(x)}{\partial x}%
_{x=L^{-}}+  \label{system2} \\
&&G^{\uparrow }\hat{u}^{\uparrow }\hat{f}^{N}(L^{-})\hat{u}^{\uparrow
}+G^{\downarrow }\hat{u}^{\downarrow }\hat{f}^{N}(L^{-})\hat{u}^{\downarrow
}+  \nonumber \\
&&G^{\uparrow \downarrow }\hat{u}^{\uparrow }\hat{f}^{N}(L^{-})\hat{u}%
^{\downarrow }+\left( G^{\uparrow \downarrow }\right) ^{\ast }\hat{u}%
^{\downarrow }\hat{f}^{N}(L^{-})\hat{u}^{\uparrow }  \nonumber \\
&=&\text{ }G^{\uparrow }\hat{u}^{\uparrow }\hat{f}_{{\cal R}}^{F}\hat{u}%
^{\uparrow }+G^{\downarrow }\hat{u}^{\downarrow }\hat{f}_{{\cal R}}^{F}\hat{u%
}^{\downarrow }.  \nonumber
\end{eqnarray}
\end{mathletters}%
%
The set of parameters $\left\{ G^{\uparrow },G^{\downarrow },%
\mathop{\rm Re}%
G^{\uparrow \downarrow },%
\mathop{\rm Im}%
G^{\uparrow \downarrow },\hat{u}^{\uparrow },\hat{u}^{\downarrow }\right\} $
is in general different for each contact, but we have omitted the indices $%
{\cal L}$ and ${\cal R}$ in (\ref{system1}) and (\ref{system2}) for brevity.
Eqs. (\ref{system1}) and (\ref{system2}) are two $2\times 2$ matrix
equations, that provide us a system of linear equations that determinate the
eight unknown constants (${\cal O},{\cal P},{\cal A},{\cal B},{\cal C},{\cal %
D},{\cal E},{\cal F}$).

From (\ref{partcurrentdens.}) we can see that the total particle current $%
i_{0}^{N}$\ can be written in terms of one of these constants as: 
\[
i_{0}^{N}=-2S\nu _{_{DOS}}D\frac{\partial f_{0}(x)}{\partial x}=-2DS\nu
_{_{DOS}}{\cal O}=-2V_{ol}G_{N}{\cal O}, 
\]
where $G_{N}=\frac{D}{L}\nu _{_{DOS}}$ is the normal metal conductance and $%
V_{ol}$ is the volume of the normal metal. By solving the system of
equations (\ref{system1}) and (\ref{system2}), we can calculate this total
particle current. $i_{0}^{N}$ is proportional to the difference between the
distribution functions of the ferromagnets $i_{0}^{N}\varpropto \left( f_{%
{\cal L}}^{F}-f_{{\cal R}}^{F}\right) $, times a quantity which does not
depends on energy. From this quantity is possible to obtain the total
conductance $G^{T}$: 
\begin{equation}
\text{ }i_{0}^{N}=G^{T}\left( f_{{\cal L}}^{F}-f_{{\cal R}}^{F}\right)
\label{CONDUCTANCE}
\end{equation}
where $G^{T}$ is in principle a function of the relative orientation of the
magnetization directions in the ferromagnetic reservoirs, the contacts and
normal metal conductances, the spin diffussion length and also of the
magnetic field:\cite{comment2}

\[
G^{T}{\bf \equiv }G^{T}\left( {\bf \vec{m}}_{{\cal L},{\cal R}},\left\{
G^{\uparrow },G^{\downarrow },%
\mathop{\rm Re}%
G^{\uparrow \downarrow },%
\mathop{\rm Im}%
G^{\uparrow \downarrow }\right\} _{{\cal L},{\cal R}},G_{N},l_{sf},\vec{B}%
\right) . 
\]
By studying $G^{T}$ for different values of these parameters, we obtain
information about the physics of the spin accumulation in diffusive systems.

\section{\bf ANALYTICAL EXPRESSIONS.}

The properties of the contacts are parametrized by the spin-dependent
conductances $\left\{ G^{\uparrow },G^{\downarrow },%
\mathop{\rm Re}%
G^{\uparrow \downarrow },%
\mathop{\rm Im}%
G^{\uparrow \downarrow }\right\} _{{\cal L},{\cal R}}$. For collinear
configurations of the ferromagnetic reservoirs (parallel and antiparallel),
it is easy to obtain simple expressions for the conductance, which can be
interpreted by simple equivalent circuits. When $l_{sf}\gg L$, there is no
mixing between spin-up ($\uparrow $) and spin-down ($\downarrow $) channels,
and we obtain the conductance for the parallel configuration

\begin{mathletters}%
%

\begin{equation}
G_{P}=\frac{G_{{\cal L}}^{\uparrow }G_{{\cal R}}^{\uparrow }G_{N}}{\left( G_{%
{\cal L}}^{\uparrow }+G_{{\cal R}}^{\uparrow }\right) G_{N}+G_{{\cal L}%
}^{\uparrow }G_{{\cal R}}^{\uparrow }}+\frac{G_{{\cal L}}^{\downarrow }G_{%
{\cal R}}^{\downarrow }G_{N}}{\left( G_{{\cal L}}^{\downarrow }+G_{{\cal R}%
}^{\downarrow }\right) G_{N}+G_{{\cal L}}^{\downarrow }G_{{\cal R}%
}^{\downarrow }}  \label{GPGeneral}
\end{equation}
and for the antiparallel configuration: 
\begin{equation}
G_{AP}=\frac{G_{{\cal L}}^{\uparrow }G_{{\cal R}}^{\downarrow }G_{N}}{\left(
G_{{\cal L}}^{\uparrow }+G_{{\cal R}}^{\downarrow }\right) G_{N}+G_{{\cal L}%
}^{\uparrow }G_{{\cal R}}^{\downarrow }}+\frac{G_{{\cal L}}^{\downarrow }G_{%
{\cal R}}^{\uparrow }G_{N}}{\left( G_{{\cal L}}^{\downarrow }+G_{{\cal R}%
}^{\uparrow }\right) G_{N}+G_{{\cal L}}^{\downarrow }G_{{\cal R}}^{\uparrow }%
}.  \label{GAPGeneral}
\end{equation}

\end{mathletters}%
%
On the other hand, when $l_{sf}\ll L$, spin-up ($\uparrow $) and spin-down ($%
\downarrow $) channels are completely mixed due to spin-flip scattering and
the spin accumulation vanishes. In this limit we have: 
\begin{equation}
G^{0}=\left( \frac{1}{2G_{N}}+\frac{1}{G_{{\cal L}}^{\uparrow }+G_{{\cal L}%
}^{\downarrow }}+\frac{1}{G_{{\cal R}}^{\uparrow }+G_{{\cal R}}^{\downarrow }%
}\right) ^{-1}.  \label{GOGeneral}
\end{equation}
These expressions correspond to the simple equivalent circuits displayed in
Fig. 3. Eq. (\ref{GPGeneral}) corresponds to a circuit in which the two spin
channels are independent in the parallel configuration (Fig. 3a). Eq. (\ref
{GAPGeneral}) corresponds to the anti-parallel configuration (Fig. 3b). Eq. (%
\ref{GOGeneral}) is equivalent to a circuit with a complete mixing between
spin up ($\uparrow $) and spin-down ($\downarrow $) channels (Fig. 3c), in
which there is no difference between parallel and anti-parallel
configurations.

For symmetric contacts $\left( G_{{\cal L}}^{\uparrow }=G_{{\cal R}%
}^{\uparrow }=G^{\uparrow }\text{ and }G_{{\cal L}}^{\downarrow }=G_{{\cal R}%
}^{\downarrow }=G^{\downarrow }\right) $, we find analytical expressions for
the conductance of the system for any value of $L/l_{sf}$, in the parallel
configuration:

\begin{mathletters}%
%
\begin{equation}
G_{P}^{S}=2G_{N}\frac{2G^{\uparrow }G^{\downarrow }\frac{l_{sf}}{L}\tanh (%
\frac{L}{2l_{sf}})+GG_{N}}{G_{N}\left( 4G_{N}+G\right) +2(G_{N}G+G^{\uparrow
}G^{\downarrow })\frac{l_{sf}}{L}\tanh (\frac{L}{2l_{sf}})}
\label{GanalitycP}
\end{equation}
and in the antiparallel configuration: 
\begin{equation}
G_{AP}^{S}=2G_{N}\frac{2G^{\uparrow }G^{\downarrow }\frac{l_{sf}}{L}%
+GG_{N}\tanh (\frac{L}{2l_{sf}})}{G_{N}\left( 4G_{N}+G\right) \tanh (\frac{L%
}{2l_{sf}})+2(G_{N}G+G^{\uparrow }G^{\downarrow })\frac{l_{sf}}{L}}
\label{GanalitycAP}
\end{equation}
\end{mathletters}%
%
where $G=G^{\uparrow }+G^{\downarrow }$.

In the limit $l_{sf}\gg L$, these equations reduce to, for parallel
configuration:

\begin{mathletters}%
%
\begin{equation}
G_{P}^{S}=\frac{G^{\uparrow }G_{N}}{G^{\uparrow }+2G_{N}}+\frac{%
G^{\downarrow }G_{N}}{G^{\downarrow }+2G_{N}},  \label{GspinP}
\end{equation}
for anti-parallel configuration: 
\begin{equation}
G_{AP}^{S}=2\frac{G^{\uparrow }G^{\downarrow }G_{N}}{GG_{N}+G^{\uparrow
}G^{\downarrow }},  \label{GspinAP}
\end{equation}
\end{mathletters}%
%
and in the limit $\left( l_{sf}\ll L\right) $ for parallel and anti-parallel
configurations: 
\begin{mathletters}%
%
\begin{equation}
G^{0}=\frac{GG_{N}}{2G_{N}+G/2}.  \label{GoAP}
\end{equation}
\end{mathletters}%
%

For non-collinear configurations there is no simple circuit analogy, but we
can still find an analytical expression for the total conductance of the
system as a function of the angle between the magnetizations of the
different ferromagnets $\theta $, when $l_{sf}\gg L$, at zero magnetic field
($\vec{B}=0$) and for symmetric contacts: 
\begin{eqnarray}
{G^{T}(\theta )=2G_{N}} \hspace{0.8\textwidth}  \label{G(alpha)} \\
\times \left( 1-\frac{\left| G^{\uparrow \downarrow }\right| ^{2}\left(
4G_{N}^{2}\left( 1+\cos \theta \right) +2GG_{N}\right) +2%
\mathop{\rm Re}%
G^{\uparrow \downarrow }G_{N}^{2}G\left( 1-\cos \theta \right) }{\left|
G^{\uparrow \downarrow }\right| ^{2}\left( (4G_{N}^{2}+GG_{N})\left( 1+\cos
\theta \right) +2\left( GG_{N}+G^{\uparrow }G^{\downarrow }\right) \right) +2%
\mathop{\rm Re}%
G^{\uparrow \downarrow }G_{N}\left( \left( GG_{N}+G^{\uparrow }G^{\downarrow
}\right) \left( 1-\cos \theta \right) \right) }\right) .  \nonumber
\end{eqnarray}

In the limit of $\theta =0$ and $\theta =\pi $, Eq. (\ref{G(alpha)})
simplifies to Eqs. (\ref{GspinP}) and (\ref{GspinAP}) respectively. When the
resistance of the normal metal is negligible compared to the contacts
resistance ($G_{N}\rightarrow \infty $), this reduces to 
\begin{equation}
G^{T}\left( \theta \right) =\frac{G}{2}\left( 1-p^{2}\frac{\tan ^{2}\theta /2%
}{\tan ^{2}\theta /2+\left| \eta \right| ^{2}/%
\mathop{\rm Re}%
\left( \eta \right) }\right)  \label{G_arne}
\end{equation}
where $p=P/G=\left( G^{\uparrow }-G^{\downarrow }\right) /G$ is the
polarization and $\eta =2G^{\uparrow \downarrow }/G$ is the (complex)
relative mixing conductance. Eq. (\ref{G_arne}) can also be obtained by
means of the circuit theory.\cite{RefYuli,RefArne}

\section{\bf NUMERICAL RESULTS.}

The total conductance depends on the spin-dependent conductances of the
contacts. We mostly set the polarization $p=P/G=\left( G^{\uparrow
}-G^{\downarrow }\right) /G=0.5$ (for real metallic ferromagnets like Fe or
Co, $p$ is 0.4 and 0.35 respectively\cite{RefTedrow}), which corresponds to
a ratio $G^{\uparrow }/G^{\downarrow }=3$. On the other hand, the real part
of the mixing conductance obeys $%
\mathop{\rm Re}%
G^{\uparrow \downarrow }\geq \left( G^{\uparrow }+G^{\downarrow }\right) /2$%
. \cite{RefCircuit} The conductances of the contacts and the diffusive
normal metal are considered to be of the same order $G_{N}\sim \left(
G^{\uparrow },G^{\downarrow },G^{\uparrow \downarrow }\right) .$

\subsection{\bf Collinear and non-collinear configurations}

The total conductance depends on the magnetic configuration. We plot in Fig.
4a and Fig. 4b, $G^{T}/G_{N}$ as function of the relative angle between
magnetizations $\theta $, for symmetric contacts, zero magnetic field ($\vec{%
B}=0$) and in the absence of spin-relaxation in the normal metal $\left(
l_{sf}\gg L\right) $, as given by Eq. (\ref{G(alpha)}) for different values
of $%
\mathop{\rm Im}%
G^{\uparrow \downarrow }$ and $%
\mathop{\rm Re}%
G^{\uparrow \downarrow }$ respectively. For $\theta =0^{%
{{}^\circ}%
},$ $\theta =360^{%
{{}^\circ}%
}$ and $\theta =180^{%
{{}^\circ}%
}$ the total conductance does not depend on the mixing conductance and the
values of $G^{T}/G_{N}$ at $\theta =0^{%
{{}^\circ}%
},360^{%
{{}^\circ}%
}$ and $\theta =$ $180^{%
{{}^\circ}%
}$ are given by Eq. (\ref{GspinP}) and Eq. (\ref{GspinAP}) respectively. On
the other hand, for non-collinear configurations, the total conductance
increases with increasing mixing conductance (the dip become more sharp).
This enhancement is due to the contributions of non-collinear spins to the
transport, in which electrons with spins oriented in different directions
than the magnetization of the adjacent ferromagnet are transmitted or
reflected at the contact. These processes are described by the real and the
imaginary part of the mixing conductance.

\subsection{\bf Spin-flip scattering. Spin relaxation}

When spin-flip scattering is caused by spin-orbit interaction in the normal
metal, the spin diffusion length $l_{sf}$ can be estimated to be equal to $%
l_{f}/(\alpha Z)^{2}$, where $\alpha $ is the relativistic fine structure
constant, $Z$ is the atomic number and $l_{f}$ is the mean free path (see 
{\em e.g.} Ref.\onlinecite{RefGorkov}). In Co/Cu multilayers, the spin
diffusion length $l_{sf}$ is of the order of a few hundred angstrom (see
Appendix A in Ref.\onlinecite{RefValet}). For Al, $l_{sf}$ can be estimated
to be of the order of a few micrometers for polycrystalline Al (see Ref.%
\onlinecite{RefTedrow}), or even between $10-70\mu m$ for Al-single crystals.
\cite{RefMJohnson} In the case of very pure Na, $\tau _{sf}\sim 1\mu s$.\cite
{RefKolbe} In this case, $l_{sf}$ limited by spin-orbit interactions can be
estimated to be of the order of $0.4$ $cm$. In Fig. 5a we plot, at zero
magnetic field and for symmetric contacts, the conductance of the system $%
G^{T}$, normalized to the conductance $G^{0}$ given by Eq. (\ref{GoAP}), as
a function of $L/l_{sf}$. The length of the normal metal section $L$ is set
to be constant, in order to keep a constant value of $G_{N}$. When $%
l_{sf}\gg L$ the conductance of the system depends on the magnetic
configuration. By decreasing $l_{sf}$, all configurations converge to the
same value of conductance $G^{T}/G^{0}=1$. All configurations reach the same
value of the conductance long before $G^{T}/G^{0}=1$, since for $l_{sf}<L$
both contacts become independent and as the relative magnetic configuration
is irrelevant. In Fig. 5b we plot $G^{T}/G^{0}$ in the case of antiparallel
configuration for differents values of the relative polarization $P/G_{N}=$ $%
\left( G^{\uparrow }-G^{\downarrow }\right) /G_{N}$ and for $G/G_{N}=\left(
G^{\uparrow }+G^{\downarrow }\right) /G_{N}$ constant. When $l_{sf}\gg L$
the configuration with large relative polarization $P/G_{N}$ gives a small
conductance and {\em vice versa}. The spin accumulation increases with
increasing polarization of the ferromagnet and causes a reduction of the
total conductance of the system. For $l_{sf}\ll L$ we also see that in each
case the conductance approaches $G^{0}$ asymptotically in different ways,
depending on the magnitude of the spin accumulation.

\subsection{\bf Effect of the magnetic field. Precession and relaxation}

In a diffusive system the presence of an external magnetic field relaxes the
spin accumulation, in addition to the usual precession of the spin.
Semiclassically, the spin accumulation at a certain position $x$ is the
average contribution of the spin of all electrons. In a diffusive metal each
electron diffuses along a random trajectory, while its spin precesses with
frequency $\omega _{L}$ around the magnetic field. Since each trajectory has
a different length, the spins of the electrons at a certain point $x$ are
oriented in different directions, which in average relaxes the local spin
polarization. The length scale of both relaxation and precession processes
is the {\em precession }length $l_{B}=$ $\sqrt{2\hslash D/g\mu _{B}B}$,
where $D$ is the diffusion coefficient, $B$ is the magnetic field, $\mu _{B}$
is the Bohr magneton and $g$ is the spin gyromagnetic ratio.

The external magnetic field may also influence the transport processes
described by the mixing conductance $G^{\uparrow \downarrow }$ at the
contacts. Let us consider for simplicity that $G_{N}\gg \left( G^{\uparrow 
\text{ }},G^{\downarrow \text{ }},G^{\uparrow \downarrow }\right) $, and $%
l_{sf}\rightarrow \infty .$ In this limit the distribution function of the
normal metal does not depend on position. From current conservation we have:

\begin{eqnarray}
i_{{\cal L}}^{\text{ }C}+i_{{\cal R}}^{\text{ }C} &=&0  \label{INVARIANT1} \\
\vec{i}_{{\cal L}}^{\text{ }C}+\vec{i}_{{\cal R}}^{\text{ }C} &=&\left( 
\frac{g\mu _{B}}{\hslash }\vec{B}\times \vec{f}\right) V_{ol}
\label{INVARIANT2}
\end{eqnarray}
where Eq. (\ref{INVARIANT1}) corresponds to the particle current, Eq. (\ref
{INVARIANT2}) corresponds to the spin current and $V_{ol}$ is the volume of
the normal metal. The current is defined to be positive when injected into
the normal metal by the ferromagnetic reservoirs. We can re-write (\ref
{INVARIANT2}) as 
\[
\vec{i}_{{\cal L}}^{\text{ }C}+\vec{i}_{{\cal R}}^{\text{ }C}=\left( g\vec{%
\omega}_{L}\times \vec{f}\right) V_{ol} 
\]
where $\vec{\omega}_{L}$ is the Larmor frequency vector. From this
expression follows that the time scale relevant for $\omega _{L}$ is the
escape time $\tau _{esc}=e^{2}$ $\nu _{_{DOS}.}V_{ol}/G^{contact}$, where $%
G^{contact}$ is the average contact conductance. $\tau _{esc}$ is the time
in which an electron escapes from the normal metal into the ferromagnetic
reservoirs. It is also the time scale relevant for the precession of the
electrons around the magnetic field. On the other hand, if $G_{N}\sim \left(
G^{\uparrow \text{ }},G^{\downarrow \text{ }},G^{\uparrow \downarrow
}\right) $, $\tau _{esc}$ is of the order of the Thouless time $\tau _{D}$,
which is the average time in which an electron passes through the diffusive
normal metal. When $G_{N}$ $\sim \left( G^{\uparrow \text{ }},G^{\downarrow 
\text{ }},G^{\uparrow \downarrow }\right) $ diffusion, precession and
transmission or reflection at the contacts, happen on the same time scale.
From these estimates we see that the ballistic or diffusive nature of the
normal metal is not going to change the effect of the magnetic field on the
physics at the contacts. The results obtained for $G_{N}\gg \left(
G^{\uparrow \text{ }},G^{\downarrow \text{ }},G^{\uparrow \downarrow
}\right) $, should therefore be valid when $G_{N}$ $\sim \left( G^{\uparrow 
\text{ }},G^{\downarrow \text{ }},G^{\uparrow \downarrow }\right) $. We now
make a perturbation expansion in small magnetic fields (see Appendix A). To
first order, the current depends on the expansion parameter $B$ as: 
\begin{equation}
\frac{\partial i_{0}}{\partial B}=\vec{s}\text{ }{\bf \hat{C}}^{-1}{\bf \hat{%
M}}\text{ }{\bf \hat{C}}^{-1}\text{ }\vec{b}.  \label{FINALEXPANS}
\end{equation}
where $\vec{s}$ and $\vec{b}$ are vectors associated with the spin current
injected into the normal metal (see Appendix A) and where the matrix ${\bf 
\hat{C}}$ describes the contacts and ${\bf \hat{M}}$ the magnetic field
contribution. As detailed in Appendix A, ${\bf \hat{C}}$ has a symmetric
part ${\bf \hat{S}}_{{\bf \hat{C}}}$, which only includes three of the four
contact conductances, {\em i.e., }only{\em \ }the conductances{\em \ }$%
G^{\uparrow },G^{\downarrow },%
\mathop{\rm Re}%
G^{\uparrow \downarrow }$ of each contact (${\cal L}$ and ${\cal R}$)
respectively. On the other hand, ${\bf \hat{C}}$ has also an antisymmetric
part which only depends on the imaginary part of the mixing conductance of
each contact $%
\mathop{\rm Im}%
G_{{\cal L}\text{, }{\cal R}}^{\uparrow \downarrow }.$ The matrix ${\bf \hat{%
M}},$ which describes the precession of spins due to the magnetic field, is
also antisymmetric. Using the symmetry properties of the matrices ${\bf \hat{%
C}}$ and ${\bf \hat{M}}$ we can determine from Eq. (\ref{FINALEXPANS}) the
symmetry properties of the total conductance of the system $G^{T}=i_{0}/(f_{%
{\cal L}}^{F}$ $-f_{{\cal R}}^{F})$ with respect to the magnetic field $B$.

When ${\bf \hat{C}}$ is a symmetric matrix, $\partial G^{T}/\partial B=0$
for small values of magnetic field. The conductance of the system is then
symmetric with respect to a change of sign of the magnetic field ($\vec{B}%
\rightarrow -\vec{B}$), {\em i.e.}, with respect to time-reversal. On the
other hand, if ${\bf \hat{C}}$ is antisymmetric, $\partial G^{T}/\partial
B\neq 0$ and we can expect asymmetric behavior of the conductance with
respect to change of sign of magnetic field.

\subsubsection{Modulation of the conductance by the magnetic field. Symmetry
with respect to time reversal}

In the following, we discuss the dependence of the conductance on the
magnetic field. We obtain $G^{T}/G^{0}$ as a function of $L/l_{B}\sim L\sqrt{%
B}$ for different magnetic configurations, $l_{sf}\rightarrow \infty $ and $%
L $ constant. In Fig. 6a we plot $G^{T}/G^{0}$ in the case of symmetric
contacts, where the magnetic field is perpendicular to both magnetization
directions $\vec{B}\cdot \vec{m}_{{\cal L},{\cal R}}=0$. In this case all
injected spins precess around the magnetic field. When $l_{B}\gg L$, the
spins injected from one ferromagnet are not strongly affected by the
magnetic field, so they travel through the normal metal and reach the other
ferromagnet without relaxation. As a result, the total conductance depends
on the relative magnetic configuration. By decreasing $l_{B}$, the spin
accumulation precesses and relaxes on the scale of $l_{B}$. Due to the
precession of spins, the conductance displays in general a non-monotonic
behavior with $L/l_{B}$. This modulation of the conductance can be
understood in terms of the ``matching'' of the spins at the contacts after
precession. According to the values of the contact conductances for the
different magnetic configurations the spins are reflected or transmitted at
the contacts depending on its orientation. Concerning the relaxation, the
configuration with more spin accumulation (in this case, the antiparallel
configuration) is the most sensitive to the magnetic field (increases faster
than the other ones), since there are more spins to be rotated by the
magnetic field in this configuration than in others. In Fig. 6a, the
conductance of $\theta =180^{%
{{}^\circ}%
}$(antiparallel) and $\theta =90^{%
{{}^\circ}%
}$configurations cross the conductance for $\theta =0^{%
{{}^\circ}%
}$(parallel) around $L/l_{B}=1$. That means that at this point the spins
accumulated in these two configurations have been reduced to the value of
spin accumulation of $\theta =0^{%
{{}^\circ}%
}$configuration. After the point $L/l_{B}=1$, the parallel configuration ($%
\theta =0^{%
{{}^\circ}%
}$) gives a smaller conductance and the antiparallel configuration ($\theta
=180^{%
{{}^\circ}%
}$) gives the highest conductance. As a result, for $L/l_{B}>1$, the
parallel configuration is more sensitive to the magnetic field than the
antiparallel configuration (now the one which increases faster). The
relaxation of spins via the precession around the magnetic field depends on
the amount of spin accumulation in the system. This non-mononotic behavior
of the conductance is specially relevant between parallel and antiparallel
configurations, because the difference between the conductance of both
configurations $G_{P}^{T}-G_{AP}^{T}$ can be modulated from positive to
negative values by the external magnetic field.

When $\vec{B}\cdot \vec{m}_{{\cal L},{\cal R}}=0$, according to Eq.(\ref
{Buvw}) in Appendix A, $\vec{B}$ has only one component $B_{3}\vec{\omega}$ $%
\sim $ $B_{3}\left( \vec{m}_{{\cal L}}\times \vec{m}_{{\cal R}}\right) $
(see Ref.\onlinecite{comment3}). Moreover, the spins are injected with
directions along $\vec{m}_{{\cal L}}$ and $\vec{m}_{{\cal R}}$, so the
precession due to the magnetic field only switches the spin directions
between $\vec{m}_{{\cal L}}$ and $\vec{m}_{{\cal R}}$. As a result, the
distribution function given by (\ref{Expansionuvw}), has only two components 
$\vec{f}=f_{1}\vec{u}+f_{2}\vec{v}.$ In this particular case, ${\bf \hat{C}}$
reduces to ${\bf \hat{S}}_{{\bf \hat{C}}}$, which is a symmetric matrix. The
same holds for the matrix ${\bf \hat{M}}${\bf , }which reduces to its $%
3\times 3$ upper box, which only includes $B_{3}$ (see Appendix A). As ${\bf 
\hat{C}}$ reduces to ${\bf \hat{S}}_{{\bf \hat{C}}},$ we expect $\partial
G^{T}/\partial B=0$. Fig. 6b shows the dependence of $G^{T}/G^{0}$ on $%
B/B_{D}$ for different magnetic configurations, where $B_{D}=2\hslash /g\mu
_{B}\tau _{_{D}}$ is the scale of magnetic fields relevant for precession in
a diffusive medium. As expected, all configurations are symmetric with
respect to a change of sign in magnetic field ($\vec{B}\rightarrow -\vec{B}$%
).

\subsubsection{Modulation of the conductance by the magnetic field.
Asymmetric properties with respect to time reversal}

Now we want to investigate the role of $%
\mathop{\rm Im}%
G^{\uparrow \downarrow }$. To this end, the magnetic field is assumed to be
oriented perpendicular to both magnetizations when the system is in
collinear configurations, and parallel to the direction of one of the
magnetizations when the system is in the $\theta =90%
{{}^\circ}%
$ configuration. According to Eq.(\ref{Buvw}) in Appendix A, for $\theta =90%
{{}^\circ}%
$, the magnetic field is along $B_{1}\vec{u}+B_{2}\vec{v},$ and as a result
from the injection and precession, there are spins in the three directions $%
\vec{f}=f_{1}\vec{u}+f_{2}\vec{v}+f_{3}\vec{\omega},$ {\em i.e.}, the
precession of spins around the magnetic field induces spins along the
perpendicular direction $\left( \vec{m}_{{\cal L}}\times \vec{m}_{{\cal R}%
}\right) $ to the injection orientations $\vec{m}_{{\cal L}}$ and $\vec{m}_{%
{\cal R}}$. ${\bf \hat{C}}$ is then an antisymmetric matrix, due to the
contributions of the terms which includes $%
\mathop{\rm Im}%
G^{\uparrow \downarrow }.$ So for $%
\mathop{\rm Im}%
G^{\uparrow \downarrow }\neq 0,$ $\partial G^{T}/\partial B\neq 0,$ which
means asymmetric behavior of the conductance with respect to time reversal.
On the other hand, if we put $%
\mathop{\rm Im}%
G^{\uparrow \downarrow }=0$, ${\bf \hat{C}}$ is symmetric and $\partial
G^{T}/\partial B=0.$

In Fig. 7a we obtain $G^{T}/G^{0}$ {\em vs }$L/l_{B}$, for the same set of
parameters as in Fig. 6a. For parallel and antiparallel configurations, the
results are not modified compared to Fig. 6a. However, for $\theta =90%
{{}^\circ}%
$ the relative conductance $G^{T}/G^{0}$ does not approach unity
asymptotically. In this configuration there are some injected spins, which
are parallel to the magnetic field and which do not precess at all. So this
part of the spin accumulation remains in the system and does not relax
irrespective of the values of the magnetic field. More interesting is the
appearence of a dip in the conductance for small values of the magnetic
field. If we repeat the calculation of $G^{T}/G^{0}$ vs $L/l_{B}$ for the
same set of parameters except for $%
\mathop{\rm Im}%
G^{\uparrow \downarrow }/G_{N}=0,$ we see that the dip disappears (Fig. 7b),
so according to our discussion, it is related with asymmetric properties of
the conductance. In Fig. 7c, we plot $G^{T}/G^{0}$ vs $B/B_{D}$. As we
expect, $\theta =90%
{{}^\circ}%
$ configuration presents asymmetric behavior respect time reversal, whereas
both parallel and antiparallel configurations remain symmetric (Fig.7c). In
particular the $\theta =90%
{{}^\circ}%
$ conductance is {\em antisymmetric} with respect to time reversal, for
small values of magnetic field.

From this discussion, we understand that the real part of the mixing
conductance describes processes at the contacts in which spins perpendicular
to the magnetization direction, are transmitted or reflected obeying
time-reversal symmetry. On the other hand, the imaginary part of the mixing
conductance describes processes in which the spins precess around the
magnetization vector of the ferromagnet. As a result of the precession, the
orientation of the spin changes. The latter processes are {\em antisymmetric}
with respect to time-reversal.

\subsubsection{Supression of the magnetic field effects by spin-flip
scattering}

Spin-flip scattering causes relaxation of the spin accumulation in the
normal metal and as a result, supression of the spin-dependent properties on
the system. Now we want to investigate how the spin-flip affect the magnetic
field effects show above. The existence of spin-flip scattering reduces $%
l_{sf}$. If $l_{sf}\gg L,$ there is no strong spin-flip scattering in the
system and it is possible to observe spin-dependent effects. On the other
hand, if $l_{sf}\ll L$, the injected spins relax very fast due to spin-flip
processes and no spin-dependent effects can be observed. In particular in
Fig. 8 we show how the dip of $\theta =90%
{{}^\circ}%
$ configuration from Figs. 7a and 7c, is suppressed by spin-flip scattering
on the system. Also by decreasing $l_{sf},$ $G^{T}/G^{0}$ increases for
constant magnetic field, to the value $1$. That simply means, that the spin
accumulation relaxes due to spin-flip scattering, as is expected.

\section{\bf DISCUSSION AND CONCLUSIONS}

In this paper, the normal metal in our F-N-F device is considered three
dimensional, but can also be two dimensional (2D), {\em e.g.}, a two
dimensional electron gas (2DEG) attached to ferromagnetic reservoirs, or
even one dimensional (1D), if the normal metal is a quantum wire or a carbon
nanotube.\cite{RefAlphenaar} In this case electron-electron interaction
should be taken into account.\cite{RefBalents} The non-magnetic material can
also be a semiconductor, as shown in recent spin-injection experiments.\cite
{RefOestreich} In the case of a 2DEG attached to metallic ferromagnets, the
large difference between the conductivities of the 2DEG and the
ferromagnetic reservoirs suppresses the spin-injection via metallic
contacts. For a significant spin-injection into the 2DEG, tunnel contacts, a
semiconductor ferromagnet or a half-metallic ferromagnet are required.\cite
{RefSchmidt}

In this paper we have shown how the spin-dependent transport through a F-N-F
double heterojunction can be described in terms of the spin dependent
conductances of the contacts $\left( G^{\uparrow \text{ }},G^{\downarrow 
\text{ }},G^{\uparrow \downarrow }\right) ,$ the magnetization direction $%
\vec{m}$ of the ferromagnetic reservoirs, and the normal metal conductance $%
G_{N}.$ The dependence of the conductance on the relative angle between the
magnetizations of the different ferromagnets is affected by the mixing
conductance $G^{\uparrow \downarrow }$. For non-collinear transport between
the ferromagnetic reservoirs, $G^{\uparrow \downarrow }=%
\mathop{\rm Re}%
G^{\uparrow \downarrow }+i%
\mathop{\rm Im}%
G^{\uparrow \downarrow }$ describes transport of spins perpendicular to the
magnetization direction of the ferromagnets. These processes enhaces the
conductance for non-collinear configurations, which may be used in
multiterminal devices for modulation of the transport properties. \cite
{RefCircuit} This modulation could be useful for future applications as
spin-dependent transistors. We find that spin injection can be symmetric and
antisymmetric with respect to time-reversal. The symmetric processes are
described by $%
\mathop{\rm Re}%
G^{\uparrow \downarrow }$ and the antisymmetric ones are described by $%
\mathop{\rm Im}%
G^{\uparrow \downarrow }$. It is interesting to observe that the
antisymmetric processes described by $%
\mathop{\rm Im}%
G^{\uparrow \downarrow }$ correspond to spin precession around the
magnetization vector of the ferromagnet which couples to an external
magnetic field.

In a diffusive system, an applied magnetic field produces both precession
and relaxation of the spin accumulation. The conductance displays a
non-monotonic behavior on the scale of the {\em precession length }$l_{B}$,
which is the distance for the precession of the spin around the magnetic
field in the normal metal. Due to this modulation, the difference between
the conductances of the parallel and antiparallel configurations $%
G_{P}^{T}-G_{AP}^{T}$ can be positive and negative as a function of the
magnetic field. A possible candidate to observe this effect is Al, which has
a large $l_{sf}$ and which can be coupled to ferromagnetic reservoirs ({\em %
e.g.}, Fe, CoFe, NiFe, Co,..) via metallic junctions or also Al$_{2}$O$_{3}$
tunnel junctions$.$ Let us estimate the values of magnetic fields for
Al-single crystal associated with the points $L/l_{B}=0.5$ and $L/l_{B}=2$
of Fig. 6a, where $G_{P}^{T}-G_{AP}^{T}$ is positive and negative
respectively. If the length of the system is $L=10\mu m,$ which is
comparable with the spin diffusion length $(L\sim l_{sf}=10-70\mu m),$\cite
{RefMJohnson} we obtain for $L/l_{B}=0.5:B^{\left( +\right) }\sim $ $0.01T$
and for $L/l_{B}=2:B^{\left( -\right) }\sim 0.1T$. If $L=1\mu m$ $(L\ll
l_{sf}),$ we obtain for $L/l_{B}=0.5:B^{\left( +\right) }\sim 0.1T$ and for $%
L/l_{B}=2:B^{\left( -\right) }\sim 1T$. In both cases we see that the values
of magnetic field in the case of Al-single crystal are quite reasonable and
also that the change from positive to negative $G_{P}^{T}-G_{AP}^{T}$\ can
be achieved by an increase of the magnetic field by one order of magnitude.
The same estimate for polycrystalline Al gives us for $\ L\sim l_{sf}=0.7\mu
m,$\cite{RefTedrow} the values of $B^{\left( +\right) }\sim 1T$ and $%
B^{\left( -\right) }\sim 16T$ respectively. These fields are much higher
than the typical switching field for a ferromagnet, so polycrystalline Al
does not appear to be a good candidate. For very pure Na, if $L\sim
l_{sf}=0.4cm,$ the corresponding values of magnetic fields are $B^{\left(
+\right) }\sim \mu T$ and $B^{\left( -\right) }\sim 10\mu T$ respectively.
This modulation of $G_{P}^{T}-G_{AP}^{T}$ by a magnetic field can also be
explored in semiconductors (SC) 2DEG, as {\em e.g.} GaAs and InAs. When $%
l_{B}\sim l_{sf}$ the following expression holds for the magnetic field
corresponding to $L/l_{B}=1:B=\frac{2\hslash }{\mu _{B}}\left( \tau _{sf}%
\text{ }g\right) ^{-1}=2.27\cdot 10^{-11}\left( \tau _{sf}\text{ }g\right)
^{-1},$ which depends on the spin-flip time $\tau _{sf}$ and on the
gyromagnetic ratio $g$ of the semiconductor material. For SC, $g$ depends
strongly on the material $\left( {\em e.g.},\text{ }%
g^{GaAs}=-0.4,g^{InAs}=15.0\right) ,$ so depending on the values of $\tau
_{sf}$, one can obtain the corresponding values of magnetic field. Kikkawa
and Awschalom\cite{RefKikkawa} report $\tau _{sf}\sim 10^{-7}s$ in n-type
GaAs system, but this values corresponds to spin lifetimes of optically
pumped carriers, and not to the usual carriers relevant for transport. The
corresponding value for the magnetic field for this case is $B^{GaAs}\sim 5$ 
$10^{-4}T$. On the other hand, we are not aware of reliable values of $\tau
_{sf}$ for transport in these systems. In conclusion, from our estimates of
the relevant values of magnetic fields, Al-single crystals with
ferromagnetic contacts are good candidates to test our predictions and
possibly lead to the discovery of other new physical phenomena of
spin-transport.

\begin{center}
{\bf ACKNOWLEDGMENTS}
\end{center}

This work is part of the research program for the ``Stichting voor
Fundamenteel Onderzoek der Materie'' (FOM).We acknowledge support from the
NEDO joint research program (NTDP-98). It is a pleasure to acknowledge
useful discussions with W. Belzig, D. Pfannkuche and Y. Tokura.

\appendix

\section{{\bf PERTURBATION} {\bf EXPANSION IN SMALL MAGNETIC FIELDS}}

Eqs. (\ref{INVARIANT1}) and (\ref{INVARIANT2}) can be written as follows: 
\begin{equation}
\left( \frac{G_{{\cal L}}}{2}+\frac{G_{{\cal R}}}{2}\right) f_{0}+\frac{P_{%
{\cal L}}}{2}\left( \vec{f}\cdot \vec{m}_{{\cal L}}\right) +\frac{P_{{\cal R}%
}}{2}\left( \vec{f}\cdot \vec{m}_{{\cal R}}\right) =\frac{G_{{\cal L}}}{2}f_{%
{\cal L}}^{F}+\frac{G_{{\cal R}}}{2}f_{{\cal R}}^{F}  \label{EQINV1}
\end{equation}

\begin{eqnarray}
&&\left( \frac{G_{{\cal L}}}{2}-%
\mathop{\rm Re}%
G_{{\cal L}}^{\uparrow \downarrow }\right) \left( \vec{f}\cdot \vec{m}_{%
{\cal L}}\right) \vec{m}_{{\cal L}}+%
\mathop{\rm Re}%
G_{{\cal L}}^{\uparrow \downarrow }\text{ }\vec{f}+%
\mathop{\rm Im}%
G_{{\cal L}}^{\uparrow \downarrow }\text{ }\left( \vec{f}\times \vec{m}_{%
{\cal L}}\right) +  \nonumber \\
&&\left( \frac{G_{{\cal R}}}{2}-%
\mathop{\rm Re}%
G_{{\cal R}}^{\uparrow \downarrow }\right) \left( \vec{f}\cdot \vec{m}_{%
{\cal R}}\right) \vec{m}_{{\cal R}}+%
\mathop{\rm Re}%
G_{{\cal R}}^{\uparrow \downarrow }\text{ }\vec{f}+%
\mathop{\rm Im}%
G_{{\cal R}}^{\uparrow \downarrow }\text{ }\left( \vec{f}\times \vec{m}_{%
{\cal R}}\right)  \label{EQINV2} \\
&=&\frac{P_{{\cal L}}}{2}\left( f_{{\cal L}}^{F}-f_{0}\right) \vec{m}_{{\cal %
L}}+\frac{P_{{\cal R}}}{2}\left( f_{{\cal R}}^{F}-f_{0}\right) \vec{m}_{%
{\cal R}}-\left( \frac{g\mu _{B}}{\hslash }\vec{B}\times \vec{f}\right)
V_{ol}.  \nonumber
\end{eqnarray}
Now we expand $\vec{f}$ into a convenient basis of the vectors $\vec{m}_{%
{\cal L}},$ $\vec{m}_{{\cal R}},$ and $\vec{m}_{{\cal L}}\times \vec{m}_{%
{\cal R}}$ as

\begin{equation}
\vec{f}=f_{1}\vec{u}+f_{2}\vec{v}+f_{3}\vec{\omega}  \label{Expansionuvw}
\end{equation}
where 
\begin{eqnarray}
\vec{u} &=&\frac{\vec{m}_{{\cal L}}+\vec{m}_{{\cal R}}}{\sqrt{2\left(
1+m\right) }}  \label{u} \\
\vec{v} &=&\frac{\vec{m}_{{\cal L}}-\vec{m}_{{\cal R}}}{\sqrt{2\left(
1+m\right) }}  \label{v} \\
\vec{\omega} &=&\frac{\vec{m}_{{\cal L}}\times \vec{m}_{{\cal R}}}{\sqrt{%
1-m^{2}}}  \label{w}
\end{eqnarray}
and where $m=\vec{m}_{{\cal L}}\cdot \vec{m}_{{\cal R}}=\cos \theta $. We
can also express the magnetic field in this basis as 
\begin{equation}
\vec{B}=B_{1}\vec{u}+B_{2}\vec{v}+B_{3}\vec{\omega}.  \label{Buvw}
\end{equation}
In terms of this expansion, we can combine (\ref{EQINV1}) and (\ref{EQINV2})
into a compact matrix form as

\begin{mathletters}%
%
\begin{equation}
\left( {\bf \hat{C}}+{\bf \hat{M}}\right) \vec{a}=\vec{b}
\label{MatrixequationInV}
\end{equation}
where 
\begin{equation}
{\bf \hat{C}}=\left( 
\begin{array}{cccc}
\begin{array}{c}
\cdot
\end{array}
& 
\begin{array}{c}
\cdot
\end{array}
& 
\begin{array}{c}
\cdot
\end{array}
& 
\begin{array}{c}
0
\end{array}
\\ 
\cdot & 
\begin{array}{c}
{\bf \hat{S}}_{{\bf \hat{C}}} \\ 
(3\times 3)
\end{array}
& \cdot & 
\begin{array}{c}
\sqrt{\frac{1-m}{2}}\left( 
\mathop{\rm Im}%
G_{{\cal L}}^{\uparrow \downarrow }-%
\mathop{\rm Im}%
G_{{\cal R}}^{\uparrow \downarrow }\right)
\end{array}
\\ 
\cdot & \cdot & \cdot & -\sqrt{\frac{1+m}{2}}\left( 
\mathop{\rm Im}%
G_{{\cal L}}^{\uparrow \downarrow }+%
\mathop{\rm Im}%
G_{{\cal R}}^{\uparrow \downarrow }\right) \\ 
\text{ \ \ }0\text{ \ \ } & -\sqrt{\frac{1-m}{2}}\left( 
\mathop{\rm Im}%
G_{{\cal L}}^{\uparrow \downarrow }-%
\mathop{\rm Im}%
G_{{\cal R}}^{\uparrow \downarrow }\right) & \sqrt{\frac{1+m}{2}}\left( 
\mathop{\rm Im}%
G_{{\cal L}}^{\uparrow \downarrow }+%
\mathop{\rm Im}%
G_{{\cal R}}^{\uparrow \downarrow }\right) & 
\mathop{\rm Re}%
G_{{\cal L}}^{\uparrow \downarrow }+%
\mathop{\rm Re}%
G_{{\cal R}}^{\uparrow \downarrow }
\end{array}
\right)  \label{Mo}
\end{equation}

\begin{equation}
{\bf \hat{S}}_{{\bf \hat{C}}}=\left( 
\begin{array}{ccc}
\frac{G_{{\cal L}}+G_{{\cal R}}}{2} & 
\begin{array}{c}
\sqrt{\frac{1+m}{2}}\frac{P_{{\cal L}}+P_{{\cal R}}}{2}
\end{array}
& \sqrt{\frac{1-m}{2}}\frac{P_{{\cal L}}-P_{{\cal R}}}{2} \\[3ex] 
\begin{array}{c}
\sqrt{\frac{1+m}{2}}\frac{P_{{\cal L}}+P_{{\cal R}}}{2}
\end{array}
& \frac{(G_{{\cal L}}+G_{{\cal R}})\left( 1+m\right) }{4}+\frac{\left( 
\mathop{\rm Re}%
G_{{\cal L}}^{\uparrow \downarrow }+%
\mathop{\rm Re}%
G_{{\cal R}}^{\uparrow \downarrow }\right) (1-m)}{2}\text{ } & 
\begin{array}{c}
\text{ }\left( \frac{G_{{\cal L}}-G_{{\cal R}}}{2}-%
\mathop{\rm Re}%
G_{{\cal L}}^{\uparrow \downarrow }+%
\mathop{\rm Re}%
G_{{\cal R}}^{\uparrow \downarrow }\right) \left( \frac{\sqrt{1-m^{2}}}{2}%
\right)
\end{array}
\\[3ex] 
\begin{array}{c}
\sqrt{\frac{1-m}{2}}\frac{P_{{\cal L}}-P_{{\cal R}}}{2}
\end{array}
& \left( \frac{G_{{\cal L}}-G_{{\cal R}}}{2}-%
\mathop{\rm Re}%
G_{{\cal L}}^{\uparrow \downarrow }+%
\mathop{\rm Re}%
G_{{\cal R}}^{\uparrow \downarrow }\right) \left( \frac{\sqrt{1-m^{2}}}{2}%
\right) & \frac{(G_{{\cal L}}+G_{{\cal R}})\left( 1-m\right) }{4}+\frac{%
\left( 
\mathop{\rm Re}%
G_{{\cal L}}^{\uparrow \downarrow }+%
\mathop{\rm Re}%
G_{{\cal R}}^{\uparrow \downarrow }\right) (1+m)}{2}
\end{array}
\right)  \label{SMo}
\end{equation}

and where 
\begin{equation}
{\bf \hat{M}}=V_{ol}\frac{g\mu _{B}}{\hslash }\left( 
\begin{array}{cccc}
0 & 0 & 0 & 0 \\ 
0 & 0 & B_{3} & -B_{2} \\ 
0 & -B_{3} & 0 & B_{1} \\ 
0 & B_{2} & -B_{1} & 0
\end{array}
\right)  \label{M1}
\end{equation}
\begin{equation}
\vec{a}=\left( 
\begin{array}{c}
f_{0} \\ 
f_{1} \\ 
f_{2} \\ 
f_{3}
\end{array}
\right)  \label{vectora}
\end{equation}
\begin{equation}
\vec{b}=\frac{1}{2}\left( 
\begin{array}{c}
G_{{\cal L}}\text{ }f_{{\cal L}}^{F}\text{ }+\text{ }G_{{\cal R}}\text{ }f_{%
{\cal R}}^{F} \\ 
\sqrt{\frac{1+m}{2}}(P_{{\cal L}}\text{ }f_{{\cal L}}^{F}\text{ }+\text{ }P_{%
{\cal R}}\text{ }f_{{\cal R}}^{F}) \\ 
\sqrt{\frac{1-m}{2}}(P_{{\cal L}}\text{ }f_{{\cal L}}^{F}\text{ }-\text{ }P_{%
{\cal R}}\text{ }f_{{\cal R}}^{F}) \\ 
0
\end{array}
\right)  \label{vectorb}
\end{equation}
\end{mathletters}%
%
By a perturbation expansion in small magnetic fields, we may study how the
magnetic field is coupled with the physics at the contacts. To zeroth order
in magnetic field we simply have 
\begin{equation}
\vec{a}^{(0)}={\bf \hat{C}}^{-1}\text{ }\vec{b}.  \label{PERTUR0}
\end{equation}
To first order

\begin{equation}
\vec{a}^{(1)}=\left( {\bf \hat{C}}^{-1}{\bf \hat{M}}\text{ }{\bf \hat{C}}%
^{-1}\right) \vec{b}.  \label{PERTUR1}
\end{equation}
The total particle current in the system is given by 
\begin{equation}
i_{0}=i_{{\cal L}}^{C}-i_{{\cal R}}^{C}=\vec{s}\cdot \vec{a}+G_{{\cal L}}%
\text{ }f_{{\cal L}}^{F}\text{ }+\text{ }G_{{\cal R}}\text{ }f_{{\cal R}}^{F}
\label{PERTURCurrent}
\end{equation}
where 
\begin{equation}
\vec{s}=\left( 
\begin{array}{c}
-G_{{\cal L}}\text{ }+\text{ }G_{{\cal R}}\text{ } \\ 
\sqrt{\frac{1+m}{2}}\left( -P_{{\cal L}}\text{ }+\text{ }P_{{\cal R}}\right)
\\ 
-\sqrt{\frac{1-m}{2}}\left( P_{{\cal L}}\text{ }+\text{ }P_{{\cal R}}\right)
\\ 
0
\end{array}
\right) .  \label{PERTURg}
\end{equation}
The dependence of the current on the expansion parameter $B$ is given by 
\[
\frac{\partial \text{ }i_{0}}{\partial B}=\vec{s}\cdot \vec{a} 
\]
which to first order reduces to 
\begin{equation}
\frac{\partial \text{ }i_{0}}{\partial B}=\vec{s}\cdot \vec{a}^{(1)}=\vec{s}%
\text{ }{\bf \hat{C}}^{-1}{\bf \hat{M}}\text{ }{\bf \hat{C}}^{-1}\text{ }%
\vec{b}.  \label{PERTURDer2}
\end{equation}

\begin{center}
{\bf Figure Captions}
\end{center}

{\bf Fig. 1: }Ferromagnetic reservoirs attached to diffusive normal metal
through arbitrary contacts. Arbitrary but fixed magnetization directions of
the ferromagnetic reservoirs and orientations of a magnetic field applied to
the normal metal are taken into account. The contacts are described by the
spin-dependent conductances $G^{\uparrow },G^{\downarrow },G^{\uparrow
\downarrow }$ and the normal metal is characterized by the normal metal
conductance $G_{N}$. The ferromagnetic reservoirs are supposed to be large
enough and in local equilibrium.

{\bf Fig. 2:} Current conservation imposes boundary conditions on the
system. The current through the contacts $\hat{\imath}^{C}(x)$ is equal to
the current into the normal metal $\hat{\imath}^{N}(x)$ at each contact. $%
\hat{\imath}^{C}(x)$ depends on the contact conductances and on the
direction of the magnetization of the adjacent ferromagnet reservoir and $%
\hat{\imath}^{N}(x)$ is the current for the normal metal.

{\bf Fig. 3:} Equivalent circuits for parallel and antiparallel
configurations in the limits $l_{sf}\gg L$ and $l_{sf}\ll L$. (a) and (b)
correspond to the parallel and anti-parallel configuration respectively,
when $l_{sf}\gg L$. In this limit, the two spin channels are independent and
there is no mixing between them. (c) parallel and anti-parallel
configurations when $l_{sf}\ll L$. In this case, there is complete mixing
between spin-up ($\uparrow $) and spin-down ($\downarrow $) channels and the
spin accumulation vanishes.

{\bf Fig. 4: Mixing conductance}: Dependence of $G^{T}/G_{N}$ on the
relative angle $\theta $ between the magnetizations of the ferromagnetic
reservoirs, for symmetric contacts, zero magnetic field and in the absence
of spin-flip scattering. (a) The following set of parameters is chosen: $%
G^{\uparrow }/G_{N}=1.0,G^{\downarrow }/G_{N}=0.3,%
\mathop{\rm Re}%
G^{\uparrow \downarrow }/G_{N}=0.7$ and $%
\mathop{\rm Im}%
G^{\uparrow \downarrow }/G_{N}$ takes values 0.0, 0.5, 1.0, 2.0 and 10.0
corresponding to the different plotted lines. (b) In this case, $G^{\uparrow
}/G_{N}=1.0,G^{\downarrow }/G_{N}=0.3,%
\mathop{\rm Im}%
G^{\uparrow \downarrow }/G_{N}=0.0$ and $%
\mathop{\rm Re}%
G^{\uparrow \downarrow }/G_{N}$ changes with values 0.7, 1.0, 2.0 and 10.0.
According to the condition $%
\mathop{\rm Re}%
G^{\uparrow \downarrow }\geq \left( G^{\uparrow }+G^{\downarrow }\right) /2,$
$%
\mathop{\rm Re}%
G^{\uparrow \downarrow }/G_{N}$ cannot be, smaller than 0.65.

{\bf Fig. 5:} {\bf Effect of spin-flip scattering on the system}: For
symmetric contacts and zero magnetic field. (a) $G^{T}$ normalized to $G^{0}$
as a function of $L/l_{sf}$, for the following set of parameters: $%
G^{\uparrow }/G_{N}=1.0,$ $G^{\downarrow }/G_{N}=0.3,$ $%
\mathop{\rm Re}%
G^{\uparrow \downarrow }/G_{N}=0.7,$ $%
\mathop{\rm Im}%
G^{\uparrow \downarrow }/G_{N}=0.0$. (b) $G^{T}/G^{0}$ versus $L/l_{sf}$ in
the case of antiparallel configuration, for different values of the relative
polarization $P/G_{N}=0.1,$ $0.7,$ $1.1$ and for $G/G_{N}=1.3$ constant.

{\bf Fig. 6: Magnetic field dependence in the absence of spin-flip scattering%
}: We consider symmetric contacts and the following set of parameters: $%
G^{\uparrow }/G_{N}=1.0,$ $G^{\downarrow }/G_{N}=0.3,$ $%
\mathop{\rm Re}%
G^{\uparrow \downarrow }/G_{N}=0.7,$ $%
\mathop{\rm Im}%
G^{\uparrow \downarrow }/G_{N}=0.5$. Moreover the magnetic field is always
perpendicular to both magnetizations directions $\vec{B}\cdot \vec{m}_{{\cal %
L},{\cal R}}=0$. (a) $G^{T}/G^{0}$ as a function of $L/l_{B}$. (b) $%
G^{T}/G^{0}$ versus $B/B_{D}$.

{\bf Fig. 7: Magnetic field dependence in the absence of spin-flip scattering%
}: We consider $\vec{B}\cdot \vec{m}_{{\cal L}}=0$, $\vec{B}\cdot \vec{m}_{%
{\cal R}}^{0%
{{}^\circ}%
,180%
{{}^\circ}%
}=0$ and $\vec{B}$ $||$ $\vec{m}_{{\cal R}}^{90%
{{}^\circ}%
}$ (or $\vec{B}\cdot \vec{m}_{{\cal R}}=0$, $\vec{B}\cdot \vec{m}_{{\cal L}%
}^{0%
{{}^\circ}%
,180%
{{}^\circ}%
}=0$ and $\ \vec{B}$ $||$ $\vec{m}_{{\cal L}}^{90%
{{}^\circ}%
}$). (a) $G^{T}/G^{0}$ vs $L/l_{B},$ for symmetric contacts and the
following set of parameters: $G^{\uparrow }/G_{N}=1.0,$ $G^{\downarrow
}/G_{N}=0.3,$ $%
\mathop{\rm Re}%
G^{\uparrow \downarrow }/G_{N}=0.7,$ $%
\mathop{\rm Im}%
G^{\uparrow \downarrow }/G_{N}=0.5$.(b) Same as (a) but for $%
\mathop{\rm Im}%
G^{\uparrow \downarrow }/G_{N}=0.0.$ (c) $G^{T}/G^{0}$ versus $B/B_{D}$, for 
$G^{\uparrow }/G_{N}=1.0,$ $G^{\downarrow }/G_{N}=0.3,$ $%
\mathop{\rm Re}%
G^{\uparrow \downarrow }/G_{N}=0.7,$ $%
\mathop{\rm Im}%
G^{\uparrow \downarrow }/G_{N}=0.5.$

{\bf Fig. 8: Magnetic field dependence and spin-flip scattering}:
Conductance for $\theta =90^{%
{{}^\circ}%
}$configuration of Fig. 7a, for different ratios $l_{sf}/L=$ 0.1, 0.3, 1, 3,
10, 100.


\begin{references}
\bibitem{comment2}  The energy dependence of $G^{T}$ is only relevant for
energy scales of the order of the Fermi energy E$_{F},$ which is much higher
than the revelant energy scale for transport, so we can consider $G^{T}$ as
energy independent.
\bibitem{comment3}  Do not confuse the vector $\vec{\omega}\sim $ $\left( 
\vec{m}_{{\cal L}}\times \vec{m}_{{\cal R}}\right) $ introduced in Appendix
A, with the Larmor frequency vector $\vec{\omega}_{L}=\mu _{B}\vec{B}%
/\hslash .$
\end{references}
\end{document}